\documentstyle[aps,prl,multicol,psfig]{revtex}

\begin{document}
\draft
\title{Effect of the In-Plane Magnetic Field on Conduction of
the Si-inversion Layer: Magnetic Field Driven Disorder}
\author{V.\ M.\ Pudalov$^a$
G.\ Brunthaler$^b$, A.\ Prinz$^b$, G.\ Bauer$^b$}
\address{$^a$ P.\ N.\ Lebedev Physics
Institute, 117924 Moscow, Russia.}
\address{$^b$ Institut f\"{u}r Halbleiterphysik,
Johannes Kepler Universt\"{a}t, Linz, Austria}
\maketitle

\begin{abstract}
We compare the effects of
temperature, disorder and parallel magnetic field on the strength
of the metallic-like temperature dependence  of the resistivity.
We found a  similarity between the effects of
disorder and parallel field: the parallel field weakens the
metallic-like conduction in high mobility samples and make it
similar to that for low-mobility samples. We found
a smooth continuous effect of the in-plane field on
conduction, without any threshold. While 
conduction remains non-activated, the parallel
 magnetic field restores
 the same  resistivity value as the high
 temperature does. This matching sets substantial
 constraints on
 the choice of the theoretical models developed to explain the
 mechanism of the metallic conduction and parallel field
 magnetoresistance in 2D carrier systems.
 We demonstrate that the data for magneto-
 and temperature dependence of the resistivity of Si-MOS samples in parallel field
 may be well described by a simple mechanism of the magnetic field dependent
 disorder.
\end{abstract}

\begin{multicols}{2}

\section{Introduction}
\vspace{-0.2in}
The apparent metallic-like  temperature dependence
of the resistivity is observed in various two-dimensional (2D)
carrier systems
\cite{MOSFET_MIT,other_MIT,noscaling,hanein_9709184,hanein_9805113,hanein_9808251,simmons_9910368,ensslin}
and remains a focus of a broad interest because it challenges the
conventional theory of the metallic conduction. The effect
manifests itself in a strong drop of the resistivity over a
limited range of temperatures, $T = (0.5-0.05)T_F$, from a high
temperature value (call it `Drude' value)  $\rho_{\rm high}=\rho(T
\approx T_F)$ to a low temperature one $\rho_{0}=\rho(T \lesssim
0.05T_F)$,  here $T_F$ is the Fermi temperature. Upon lowering the
temperature further, a strong `metallic' drop in $\rho$ was found
to cross over to the conventional weak localization type
dependence \cite{gmax,hamilton_9808108,weakloc,amp,app}.
Recently, it was demonstrated \cite{weakloc,app} that the
`metallic' drop is not related to quantum interference and in
this sense should have a semiclassical origin.

Magnetic field applied in the plane of the
2D system causes a dramatic increase in resistivity
\cite{instability,simonian97,breakdown,okamoto,yoon_9907128}.
It was proven experimentally \cite{vitkalov_0101196,anisotropy}
that the  magnetoresistance (MR)
in Si-MOS samples is mainly caused by spin-effects, though certain
contribution of orbital effects \cite{hwang} is noticeable at very large fields \cite{anisotropy},
bigger than the field of the spin polarization. The
two effects, {\em strong metallic drop in
resistivity}\ \ and {\em parallel field
magnetoresistance} remain
{\em puzzling}.

In this paper we demonstrate that (i) the strength of both effects
is well matched and (ii) the action of  magnetic
field on conduction is similar to that of  disorder
and to some extent to that of  temperature. In
particular, in the presence of  in-plane magnetic
field the strong temperature dependence of the resistivity for a
high mobility sample transforms into a weaker one and shifts to
higher temperatures; both these features are typical for low
mobility samples. Next, (iii) the effect of the magnetic field on
the resistivity  is continuous and shows no signatures of a
threshold. Finally, (iv) increase in either temperature or
parallel magnetic field restores the same high-temperature
(`Drude') resistivity value while transport remains non-activated.
Matching of the actions of the above controlling parameters sets
substantial constraints on the choice of theoretical models.

We performed measurements on a number of different mobility (100)
$n-$Si-MOS samples, because this system demonstrates the most
dramatic appearance of the discussed effects. Table~\ref{samples}
below shows the relevant parameters for the four most
intensively studied samples whose mobilities
differ by a factor of\ 25; parameters for other samples
were reported in Refs. \cite{noscaling,gmax,app,pud_Hall}.
\vspace{-0.1in}
\begin{center}
\begin{minipage}{3.2in}
\begin{table}
\caption{Relevant parameters of the five samples.
$\mu_{peak}$ [m$^2$/Vs] is the peak mobility at $T =0.3$\,K,
$\rho_c$
is in units of $h/e^2$, and\  $n_c$ is in  $10^{11}$\,cm$^{-2}$.}
\begin{tabular}{|c|c|c|c|}
sample & $\mu_{\rm peak}$ & $n_{\rm c}$ & $\rho_{\rm c}$ \\
\hline
Si\ 9  & 4.3 & 0.75  & 3.2 \\
Si\ 15 & 4.0 & 0.88 & 2  \\
Si\ 43 &  1.96 & 1.4 & 0.67    \\
Si\ 39 & 0.51 & 3.5 &  0.32 \\
Si\ 46 & 0.15  & 8.3  & 0.3
\end{tabular}
\label{samples}
\end{table}
\end{minipage}
\end{center}
\vspace{-0.1in}

In Figures~1\,a,\,c  we plotted the temperature
dependence of resistivity $\rho$ for
the highest and lowest mobility samples. At high carrier density
(the lower part of Fig.~1\,a) the resistivity decreases and
demonstrates the metallic-like behaviour. The magnitude of the
drop in resistivity, $(\rho_{\rm high}-\rho_0)/\rho_0$, obviously
depends on the disorder and varies from  a factor of 5 (the lowest
curves in Fig. 1\,a) to a few \% in Fig. 1\,c. As density
decreases below a ``critical value'' $n_c$, the character of the
temperature dependence changes to the insulating one. The changes
in the character of $\rho(T)$ at $n=n_c$ are reminiscent of
 a typical metal-insulator transition (MIT) in 3D systems which is not
 expected to occur in 2D system.
In agreement with earlier
observations \cite{noscaling,hanein_9805113,yoon_9907128} the
increase in disorder causes the drop in resistivity to weaken,
the `critical' density $n_c$ to increase  and the `critical'
resistivity $\rho_c$ to decrease.

\begin{figure}
\centerline{
\psfig{figure=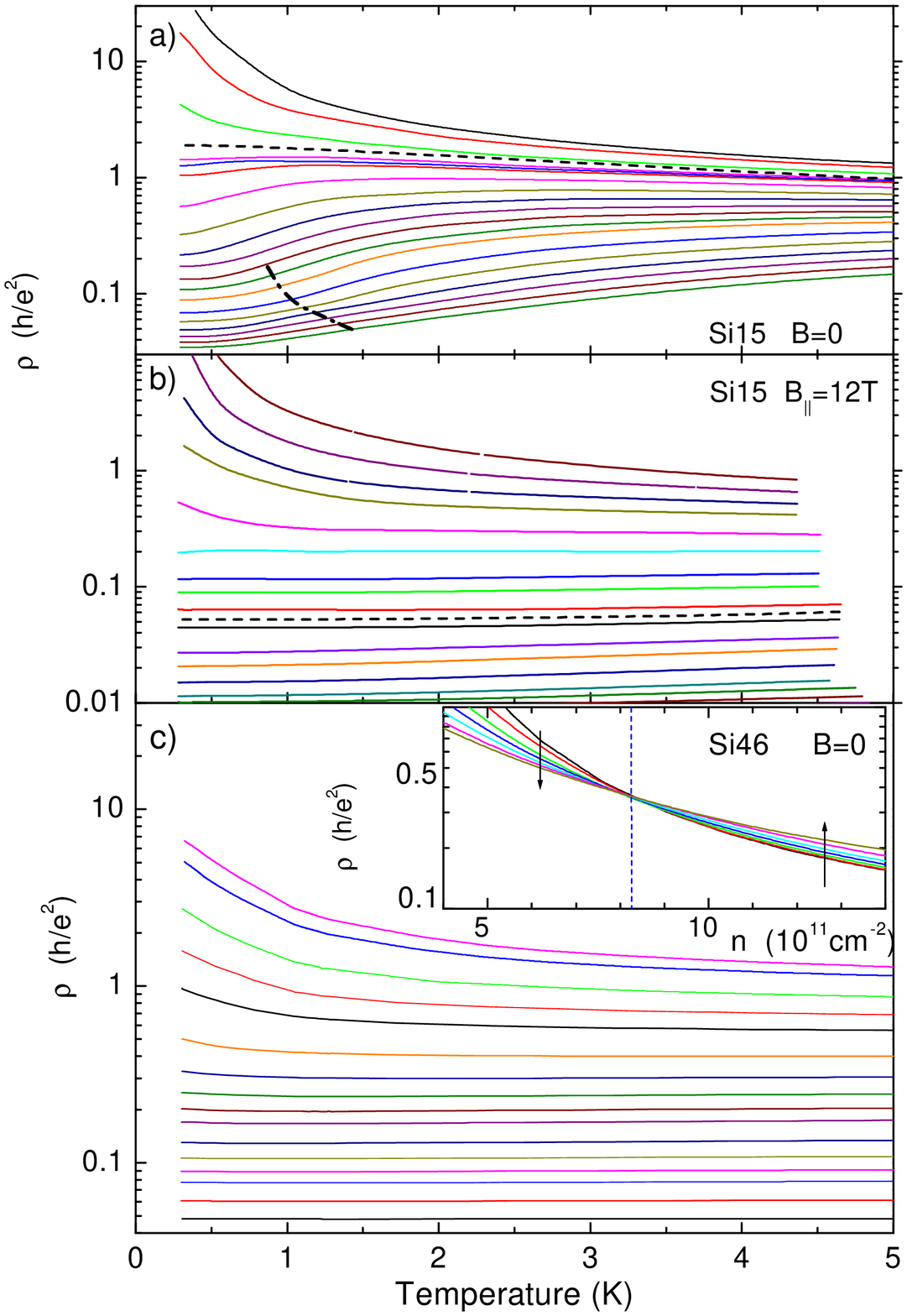,width=240pt}
}
\begin{minipage}{3.2in}
\vspace{0.1in}
\caption{$T-$dependence of the resistivity
(a) for  high mobility sample Si15 at $B=0$,  and (b)
at  $B_{\parallel}=12$\,T; (c) for
low-mobility sample Si46 at $B=0$. Inset shows  $\rho(n)$
for $T=1, 2, 4, 8, 16, 32, 62$\,K
(arrows indicate
direction of the growth in temperature).
Horizontal dashed lines correspond to a
`critical density' $n_c$.
Dash-dotted line in the panel (a) marks $T=0.1T_F$.
The densities (in units of $10^{11}$\,cm$^{-2}$)
on the panel (a) are 0,76, 0.79, 0.85, 0.89, 0.90, 0.91, 0.96,
1.01, 1.07, 1.12, 1.18, 1.23, 1.29, 1.40, 1.51, 1.62, 1.73, 1.84, 1.95;
(b)  0.95, 1.07, 1.18, 1.29, 1.51, 1.73, 2.06, 2.28, 2.61,
2.94, 3.39, 3.72; and on the panel (c) 3.85, 4.13, 4.83, 5.53,
6.23, 7.63, 9.03, 10.4, 11.8, 13.2, 16.0, 18.8, 21.6, 24.4,
30.0, 37.0.}
\label{fig1}
\end{minipage}
\end{figure}
\vspace{-0.1in}

However, in contrast  to the common believe that only high
mobility samples are subject to the MIT, we
demonstrate in the inset to Fig.~1\,c that even for such low
mobility sample as Si46, there is a clear crossing point,
$n=n_c$,  where
$d\rho(T)/dT$  changes  sign. The
difference from that for high mobility samples (Fig.~1\,a) is the
shift of the events to much higher temperatures,  to the range
$T=1-60K$, whereas for lower temperatures $T<1$\,K  the
weak localization upturn in the resistivity sets in. This upturn
is a subject of more detailed consideration \cite{app} and is not
discussed here. The increase in the temperature scale by about 10
times as a function of disorder is anticipated since the
Fermi energy, $E_F$, at the critical
carrier density for disordered sample Si46 are
10 times bigger than that for Si15; correspondingly, in both
samples the MIT is observed in similar
ranges  of $(T/E_F)$.
Comparing Figs.~1\,a and 1\,c we arrive at the conclusion that
the overall temperature dependence of  the resistivity for Si-MOS
samples may be roughly scaled by taking into account disorder
dependence of the critical density $n_c$ and the corresponding
energy scale. We ignore at the moment a {\em disorder dependence
of the strength in the resistivity drop}, which will be
considered further.
\vspace{-0.1in}
\section{The effect of the parallel magnetic field on conduction}
\vspace{-0.1in}
\subsection{Magnetic field driven MI-transition?}
\vspace{-0.1in}
It  was found in
Refs.~\cite{okamoto,yoon_9907128,simmons98} that for a fixed
carrier density an increase in $B_{\parallel}$ causes
$d\rho/dT$ to change sign.
At first sight, this effect is reminiscent of a magnetic field driven transition;
an analogy to the superconductor-insulator transition
was  discussed in Ref. \cite{NAS}
and the corresponding field value is often called
``critical magnetic field'' $B_c$.
In Figure~2 we reproduce similar behavior for $n-$Si-MOS samples.

\begin{figure}
\centerline{
\psfig{figure=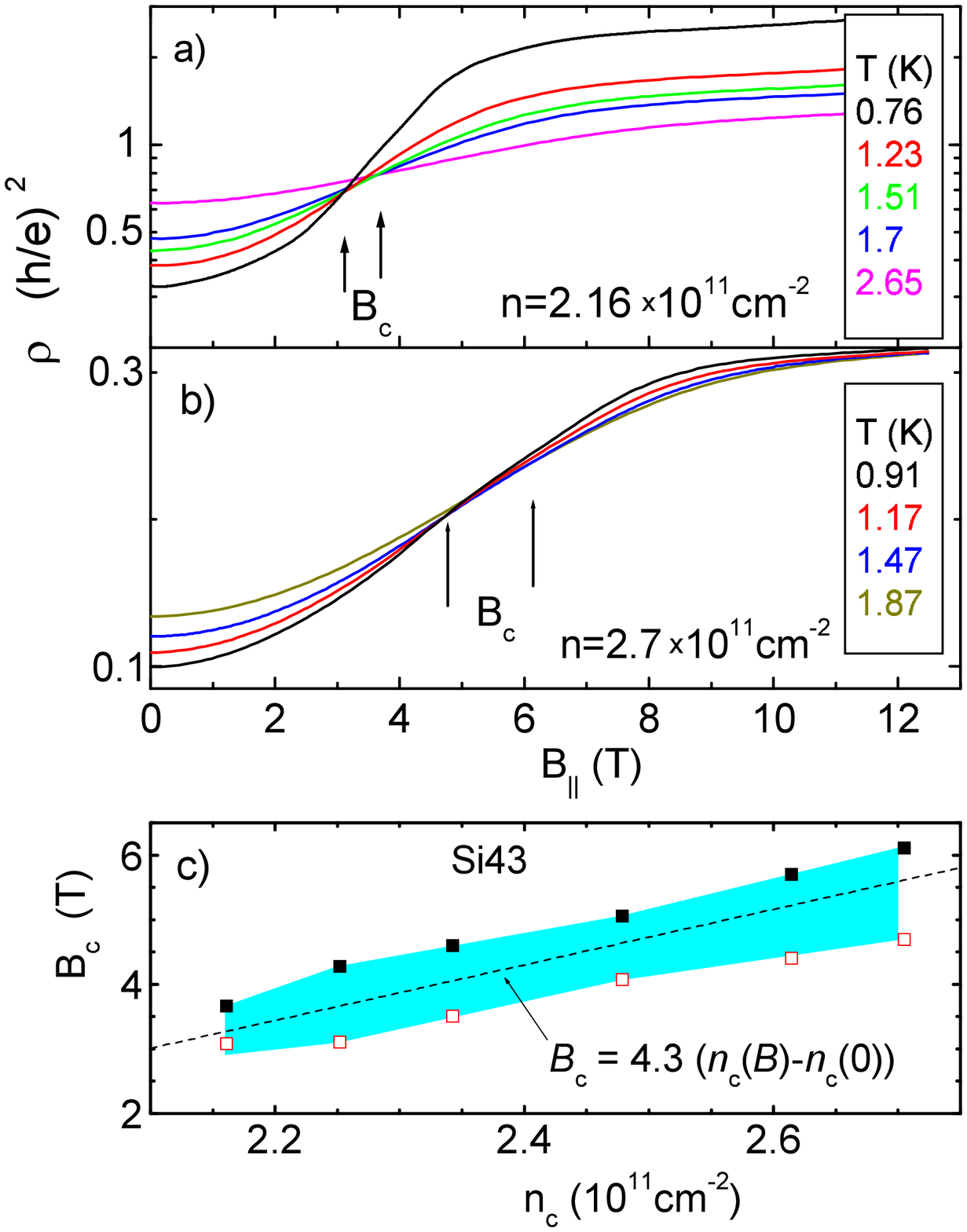,width=230pt,height=230pt}
}
\begin{minipage}{3.2in}
\vspace{0.05in}
\caption{a, b) Resistivity traces vs
$B_{\parallel}$ for different temperatures and for two densities.
Sample Si43. Vertical arrows mark interval of fields  which may be
assigned to the `critical' field value $B_c$.
c) $B_c$ vs
carrier density. Shadowed corridor corresponds to the
interval of crossing. Dashed line is a linear fit.
}
\label{fig2}
\end{minipage}
\end{figure}
\vspace{-0.1in}

Similar to that
reported in Refs. \cite{yoon_9907128,simmons98,tutuc}, the $\rho(B)$
traces for different temperatures cross each other around a `critical'
magnetic field, $B_c$,  though in an extended interval
rather than in a single point.
As carrier density $n$ and
magnetic field increase, the interval of crossing broadens and
finally becomes infinite for $B\geq 6$\,T. Density  dependence
of  $B_c$ is plotted in Fig.~2\,c in the range of fields
where the crossing interval is finite.

In order to understand the
driving mechanism of the magnetoresistance we performed more
detailed  studies of the magnetoresistance and measured its
temperature dependence at various values of the in-plane
field.

\subsection{Analogy between the effects of parallel field and disorder
on conduction}
It is rather instructive to trace in detail the influence of the
magnetic field on the density and temperature dependences of the
resistivity.
The most striking and known result of the in-plane magnetic
field is the increase
in the sample resistivity. Figure 3 demonstrates that the mobility
degrades in magnetic field smoothly, without a threshold. Besides
decreasing sample mobility, the magnetic field causes
its maximum to shift to higher density. Both these effects are
typical for the $B=0$ case when the mobility is varied by
changing the disorder; at low carrier density this is usually
described in terms of the increase in the number of scatterers,
$n_i \propto 1/\mu^{\rm peak}$ \cite{gold85}.
\vspace{0.1in}
\begin{figure}
\centerline{
\psfig{figure=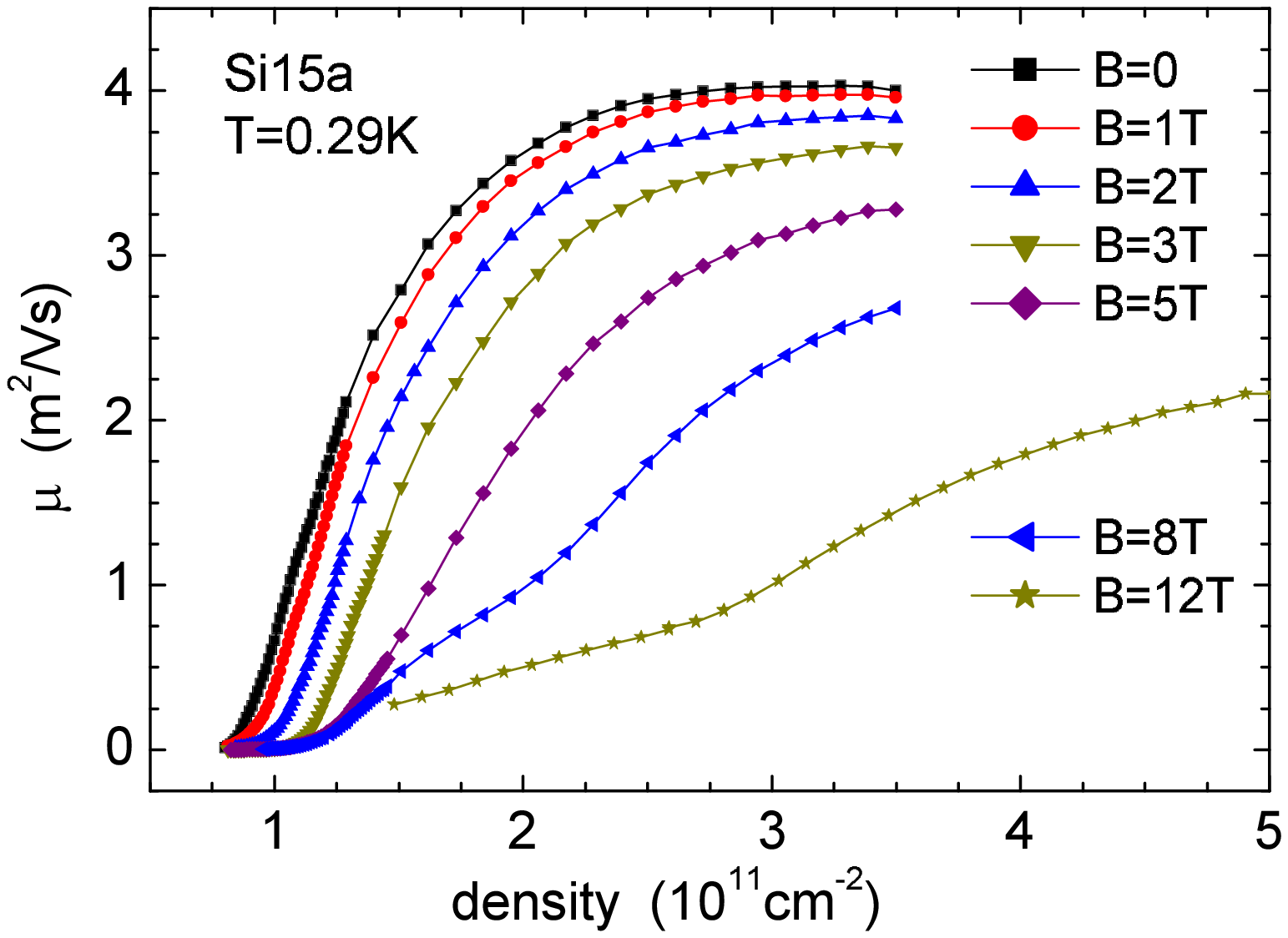,width=230pt,height=190pt}
}
\begin{minipage}{3.2in}
\vspace{0.1in} \caption{Mobility vs carrier density for
a fixed temperature $T=0.29$\,K and for different magnetic fields
$B_{\parallel}$ (indicated in the figure).}
\label{fig3}
\end{minipage}
\end{figure}
\vspace{0.1in}

In Figures 4, we show the resistivity traces vs density
for four fixed temperatures in the range  0.3 to 1\,K.
All four groups of curves for different $B_{\parallel}$ values
demonstrate a critical behaviour.
The well defined crossing point or `critical' density, $n_c$,
separates regimes
of temperature activated (i.e. `insulating', $n<n_c$)
and non-activated (`metallic', $n>n_c$) conduction \cite{definition}.

The figures clearly demonstrate again that the parallel field
progressively increases the critical carrier density $n_c$ and
decreases the corresponding critical resistivity value $\rho_c$.
For magnetic fields $B_{\parallel}>6$\,T (not shown), the
crossing point spreads into a density interval, though still
shifting to higher densities. This is
typical {\em for low
mobility samples and typical for
the effect of the disorder}.
\vspace{0.1in}
\begin{figure}
\centerline{
\psfig{figure=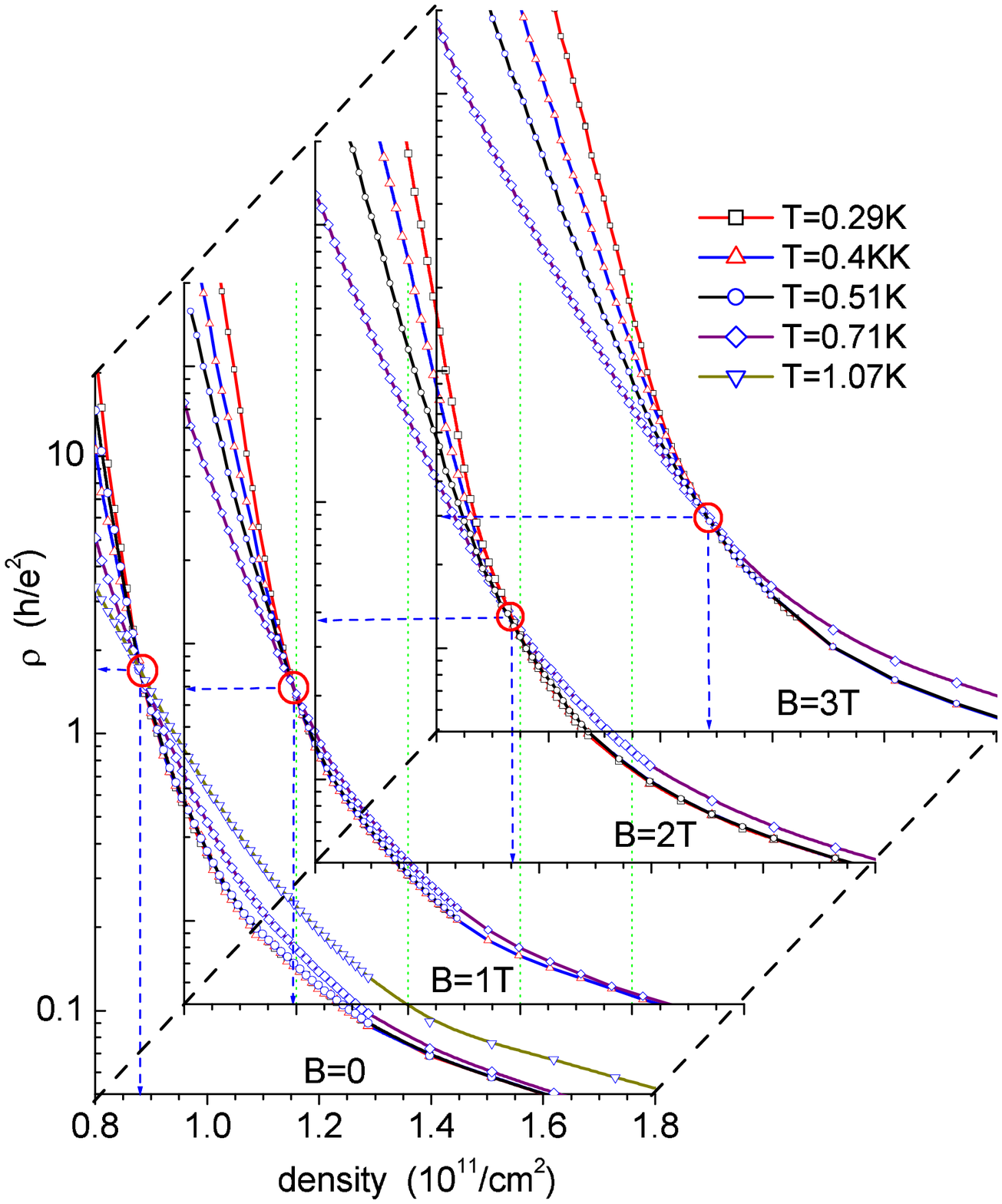,width=240pt,height=300pt}
}
\begin{minipage}{3.2in}
\vspace{0.1in}
\caption{
Resistivity vs carrier density  for 4 fixed temperatures
(indicated on the figure) measured at $B_{\parallel} =0, 1, 2$ and 3 Tesla.
Arrows at each panel mark the critical density and resistivity.}
\label{fig4}
\end{minipage}
\end{figure}

From Figure 4, we determined the coordinates of the crossing
point, $n_c$ and $\rho_c$, for each  $B_{\parallel}$ value and
plotted them in Figs.~5. These two curves $n_c(B)$ and
$\rho_c(B)$ have a transparent meaning: they separate the regimes
of the temperature activated and
`non-activated' conduction on the planes
$\rho(B,n)$ \cite{definition}.
We find again that both field dependences are
qualitatively similar to the ones known for the action of
disorder (e.g., inverse sample peak mobility
$1/\mu$, or $1/k_Fl$ value)
\cite{noscaling,hanein_9805113,yoon_9907128}.

\begin{figure}
\centerline{
\psfig{figure=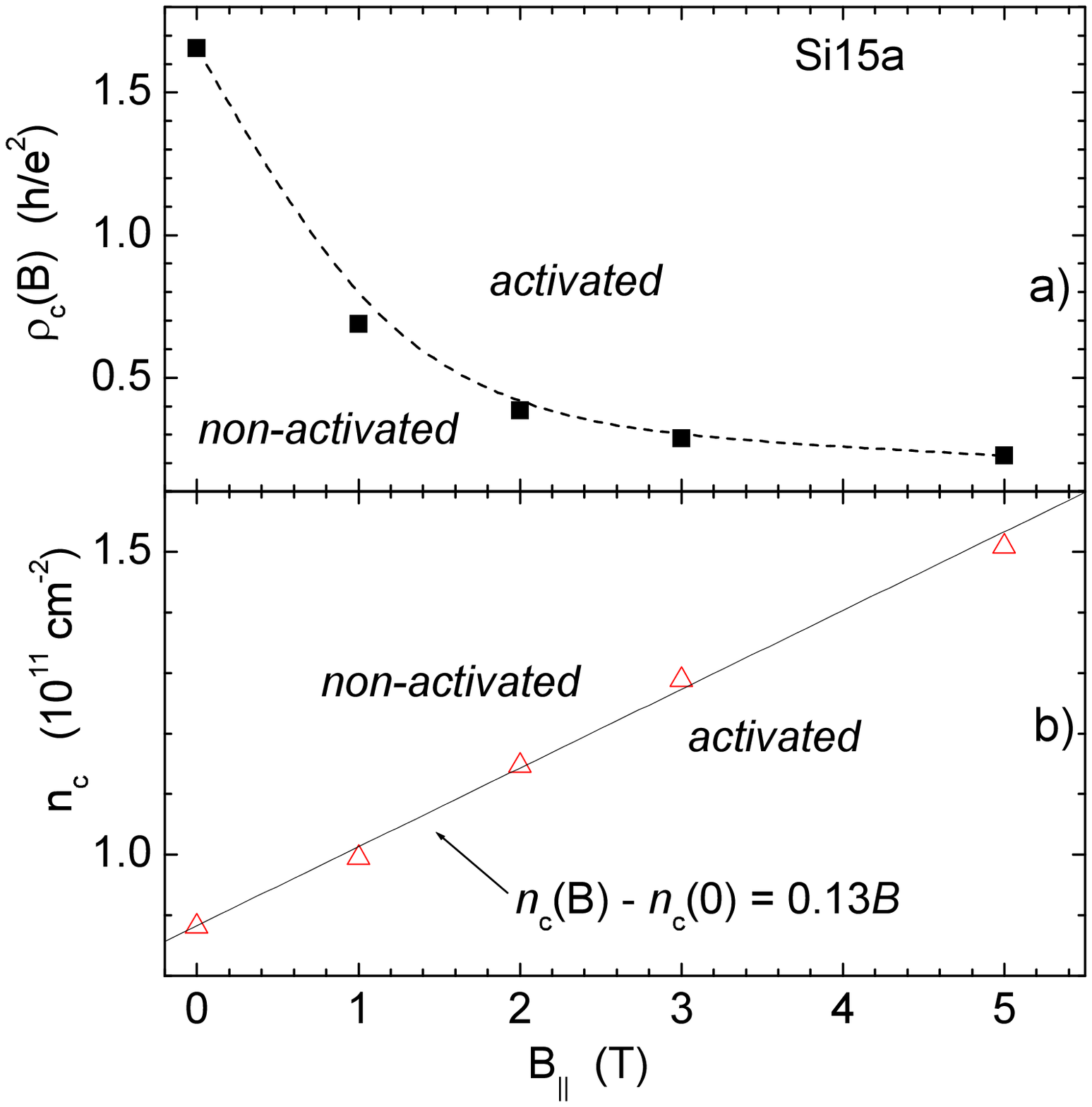,width=220pt}
}
\begin{minipage}{3.2in}
\vspace{0.1in}
\caption{Critical resistivity (a) and critical carrier density (b)
vs parallel magnetic field. Full line is a linear fit  with
adjustable slope and with $n_c=0.883 \times 10^{11}$\,cm$^{-2}$.
Dashed line is a guide to the eye.}
\label{fig5}
\end{minipage}
\end{figure}

Figure~6 shows for comparison the $\rho_c(n_c)$ dependence
measured with the sample Si15 in various parallel fields
(calculated from Fig.~4) and $\rho_c(n_c)$
  reproduced from Ref.~\cite{noscaling}
for different samples at $B=0$. A clear
analogy is seen between the decrease in $\rho_c$
caused by disorder and magnetic field.

\begin{figure}
\centerline{
\psfig{figure=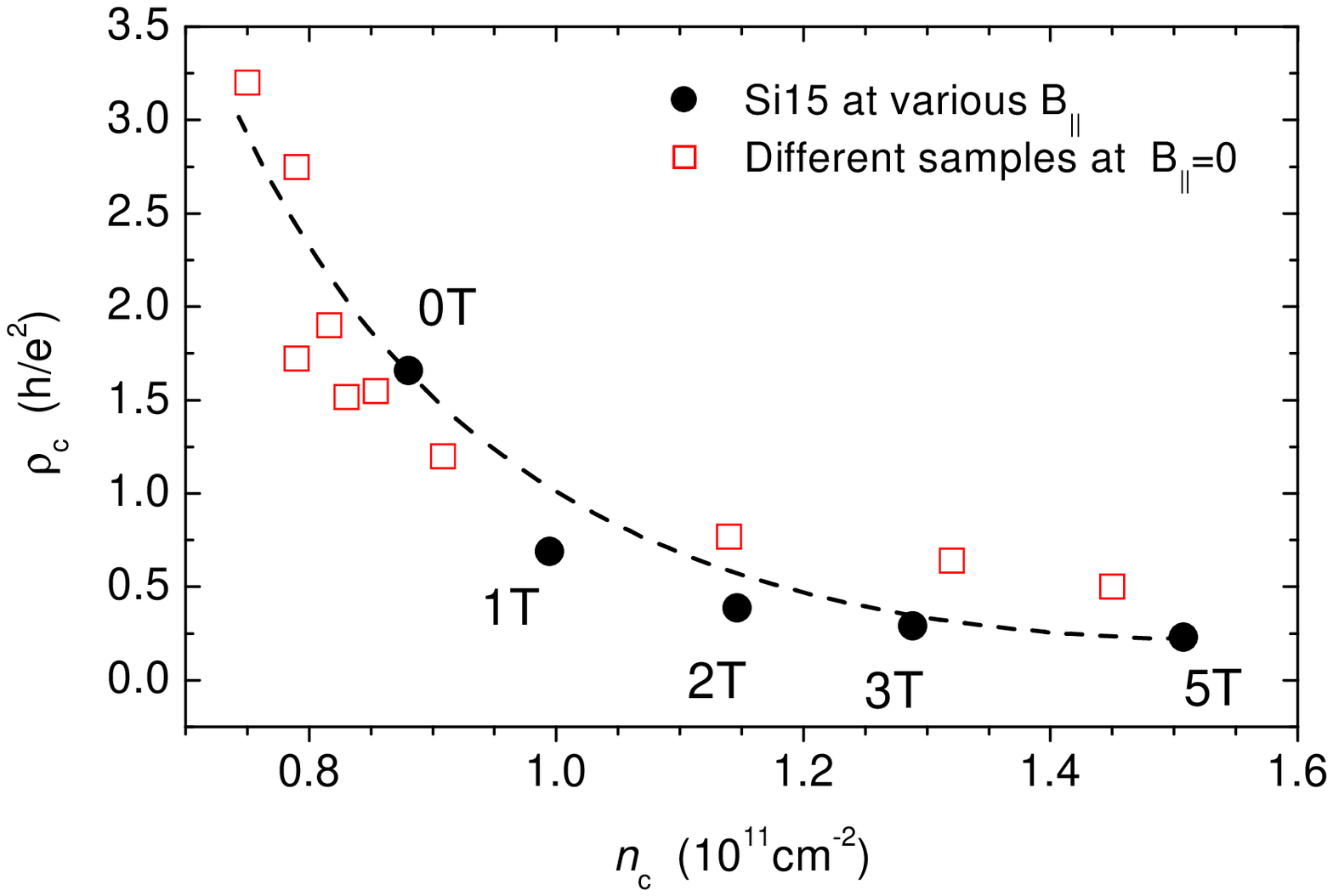,width=240pt,height=170pt}
}
\begin{minipage}{3.2in}
\vspace{0.1in}
\caption{``Critical'' resistivity value vs ``critical'' density.
Closed squares show the dependence driven by magnetic field,
open squares are for the one driven by  disorder
(plotted for 9 different samples, based on the data from
Ref.~\protect\cite{noscaling}).
Line is the  guide to the eye.}
\label{fig6}
\end{minipage}
\end{figure}

Finally, in order to illustrate the conclusion that the
action of the magnetic field is similar to that of the disorder,
we plotted in Fig.~1\,b, for comparison, $\rho(T)$ for the same high
mobility sample Si15 but measured with applied parallel field of
12\,T. In light of the above discussion, the
transformation of Fig.~1\,a into Fig.~1\,b can thus be
treated as a result of a progressive increase in electron
scattering introduced by the parallel magnetic field,
in steps like those shown in Figs.~4 and 3.
The similarity between Figs.~1\,b and 1\,c supports this
conclusion.

\subsection{Magnetoresistance in parallel field and the $g$-factor}

\begin{figure}
\centerline{
\psfig{figure=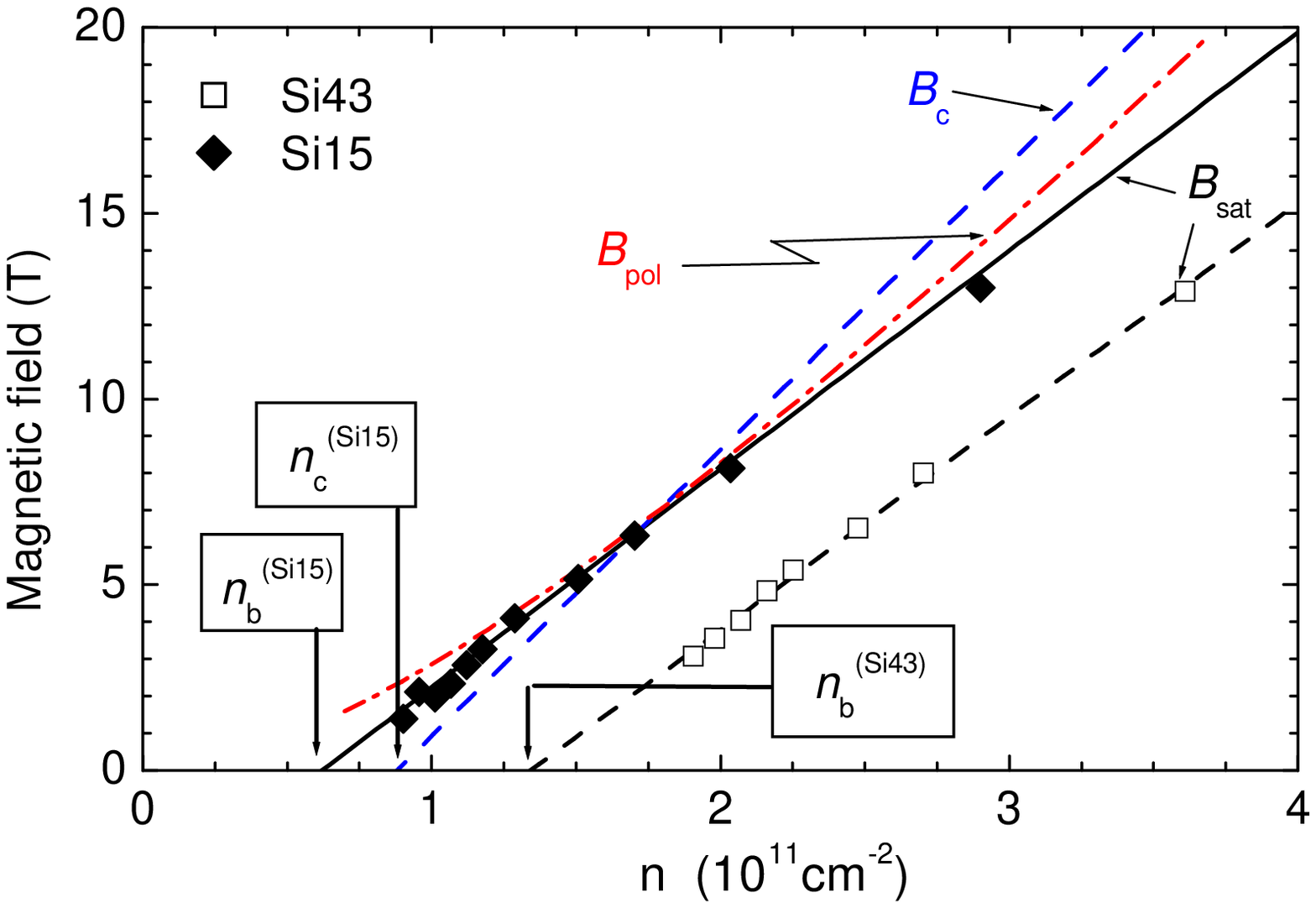,width=230pt,height=190pt,angle=0}
}
\begin{minipage}{3.2in}
\vspace{0.1in}
\caption{$B_c$ and $B_{\rm sat}$   for the sample Si15 (full diamonds)
and $B_{\rm sat}$ for Si43 (empty squares)
measured in the range $B_{\parallel}=1.3-13$\,T and
extrapolated schematically over a wider range of $B_{\parallel}$.
$B_{pol}$ is the field for complete spin polarization of the {\em conducting electrons},
calculated according to Eq.~(\protect\ref{spin_polarization}) and
using  $g^*m^*(n)$ values measured in Ref.~\protect\cite{gm}.
}
\label{B_c&B_sat}
\end{minipage}
\end{figure}

A number of experimental studies (e.g., Refs.~\cite{vitkalov_0101196,anisotropy})
demonstrate  that the parallel field magnetoresistance
in Si-MOS samples  is a spin rather than orbital effect. There were, consequently,
a number of empirical attempts to extract  the $g^*$-factor value from
different features of the magnetoresistance $R(B_{\parallel})$. For comparison,
we present  in Fig.~\ref{B_c&B_sat} the  density dependences of  two characteristic
magnetic fields:
$B_{\rm sat}$, corresponding to the saturation of $R(B_{\parallel})$ (for
Si15 and Si43) from Ref.~\cite{anisotropy}, and the `critical magnetic field' $B_c$
(for Si15) re-plotted from Fig.~5.

First of all, it is clear, that $B_{\rm sat}$ does not dependent solely on the carrier density, but
does also depend on the disorder.
Whereas the empirical dependences $B_{\rm sat} = s(n-n_b)$
for different samples have almost the same slope $s$,
the offset $n_b$ was found to be inversely proportional to the sample mobility \cite{anisotropy}.
For different samples,
$n_b$ varies from  $0.5n_c$ for the  highest mobility sample Si9  to $0.7n_c$ for Si15,
$0.8n_c$ for Si12, and $0.96 n_c$ for Si43.
On the other hand, the `critical field'
$B_c$,  by definition, extrapolates to zero at $n=n_c(B_c)$.
For  this reason, the two curves, $B_{\rm sat}(n)$ and $B_c(n)$  usually intersect.
Obviously, $n_b$ cannot be identified with $n_c$.
At low  densities, $B_{\rm sat}$ exceeds $B_c$ and the saturation of the magnetoresistance
takes place in the insulating regime. At higher densities, the saturation
occurs in the `metallic' regime;
the borderline for the sample Si15 corresponds to $n=1.75\times 10^{11}$\,cm$^{-2}$.

For comparison,
the dashed-dotted line in Fig.~\ref{B_c&B_sat} represents the calculated density dependence
of the field corresponding to the complete {\em spin polarization of conducting electrons}:
\begin{equation}
B_{\rm pol}= \frac{2E_F}{g^*\mu_B}=\left(\frac{h}{e}\right)\left(\frac{n}{g^*m^*}\right)\frac{2}{g_v},
\label{spin_polarization}
\end{equation}
where the renormalized $g^*m^*$-values for conducting electrons
were directly measured in Ref.~\cite{gm} as a function of the carrier density.

For a given density $n$, $B_{\rm sat}(n)$ is  usually {\em less} than
the field of complete spin-polarization $B_{\rm pol}$ of the conducting electrons.
The deficit, $B_{\rm pol}-B_{\rm sat}$,  increases  as sample mobility
decreases \cite{anisotropy}; for  example,
 $B_{\rm pol}-B_{\rm sat} \approx 4$\,T  for Si43.
Whereas $B_{\rm pol}$ is solely determined
by the mobile carrier density,  $B_{\rm sat}$ turns out to be
also dependent on disorder.

For high mobility samples, over a certain range of densities
(which is $(1.2-2.2)\times 10^{11}$cm$^{-2}$ for  Si15)
the two quantities, $B_{\rm pol}$ and $B_{\rm sat}$,
are rather close to each other.  This coincidence is in a good agreement with
the observation by Vitkalov et al.
 \cite{vitkalov_0101196} who  found that in the range of densities
$n=(1.54-4.5)\times 10^{11}$\,cm$^{-2}$,
the frequency of the Shubnikov-de Haas
oscillations in tilted fields  doubles at  a field  which is equal to
$B_{\rm sat}$ to within 5\%.
However, for more disordered samples (such as e.g. Si43 in
Fig.~\ref{B_c&B_sat}),  the saturation occurs
in magnetic fields  essentially lower than the polarization field $B_{\rm pol}$,
because of  the twice as large $n_b$ value (data on $n_b$ for a number of
samples may be found in Ref.~\cite{anisotropy}).
We believe therefore that the coincidence of $B_{\rm sat}$ and $B_{\rm pol}$ for
some high mobility samples and in a limited density range is rather occasional.

In the analysis of the $B_{\rm sat}(n)$-data in Ref.~\cite{vitkalov_divergence},
it was assumed that
$B_{\rm sat}$ remains equal
to $B_{pol}$ in the $n\rightarrow n_c$ limit.
With this assumption authors arrived at
the conclusion  that $g^*$ diverges as the density decreases,
and that a ferromagnetic transition takes place at the MIT. As Fig.~7 shows, the coincidence
 of $B_{\rm sat}$ and $B_{pol}$ fails for densities lower than $1.3\times 10^{11}$cm$^{-2}$
(even for the high mobility sample). The
 $g^*$-factor  for conducting electrons
 measured in Ref.~\cite{gm} from Shubnikov-de Haas effect does not diverge at $n_c$,
but gradually grows as the density decreases;
the growth is anticipated within the Fermi-liquid theory \cite{abrikosov_87,isihara_97}.

One might conclude that the non-zero value of $n_b$ is the only reason for
the difference between  $B_{\rm sat}$  and  $B_{\rm pol}$,
and that by subtracting $n_b$ the two parameters may be brought into agreement.
However, such procedure performed in  Ref.~\cite{anisotropy} for different samples,
has revealed that the two parameters are indeed
related to different physics: the $(gm)_{\rm MR}$ values
estimated from the magnetoresistance either as
$(h/e)n/B_{\rm sat}$, or $(h/e)(n-n_b)/B_{\rm sat}$
disagree with the $(g^*m^*)_{\rm SdH}$-values measured from Shubnikov-de Haas effect.
This is also seen in Fig.~\ref{B_c&B_sat}
where (i) at low densities the difference between the measured $B_{\rm sat}(n)$ values and
the $B_{pol}(n)$ curve can not be eliminated by a horizontal shift
(i.e., by varying $n_b$) and (ii) the slopes of the $B_{\rm sat}(n)$
and $B_{pol}(n)$ curves are different.

We now consider another characteristic field, $B_c$.
In Ref.~\cite{tutuc}, the $B_c(n_c)$--dependence for 2D holes
in (311)GaAs was noted to
correspond to such spin alignment in the
2D carrier system where the spin-minority population
drops below a threshold (which is approximately independent of the total 2D hole density
or the magnetic field);
the latter is of the order of
the critical density at zero field $n_c(B=0)$
\cite{tutuc}. In order to verify such
possibility for
Si-MOS samples, we fitted in Fig.~5\,b the $n_c(B)$ data for the high mobility sample
 Si15  with a linear
dependence $n_c(B)-n_b= c B$, with two adjustable parameters $n_b$ and $c$.
We find  that the offset, $n_b$,  is indeed equal
 to the anticipated value $n_c(B=0)=n_c$  (for sample parameters, see Table 1),
however, the  slope $c=0.13\times 10^{11}$cm$^{-2}$/T, disagrees with such interpretation.
With this interpretation, the slope
$c = dn_c/dB_c$ would be equal to $(g^*m^*) g_v e/2h$
(where  $g_v=2$ for $n$-(100)-Si); we obtain then
$(1/2)(g^*m^*)_{\rm MR} = 0.27$
over the displayed range of densities $n=(0.9 -
1.5)\times 10^{11}$\,cm$^{-2}$.
We repeated the same procedure with the data from
Fig.~2\,b and found, correspondingly,
$(1/2)(g^*m^*)_{\rm MR} \approx 0.48$
for the range  $n = (2.2 - 2.7)\times 10^{11}$\,cm$^{-2}$.

The increase in the slope, $dB_c/dn$  with carrier
density would mean the decrease of
$g^*m^*$  for lower densities.  Such decrease
contradicts
the results of direct measurements of the $(g^*m^*)_{\rm SdH}$ for
{\em conducting electrons}  from Shubnikov-de Haas effect
\cite{okamoto,gm};
the contradiction indicates that the
slope of the $B_c(n_c)-$dependence  in Si-MOS samples might have
a more complex interpretation \cite{anisotropy,gm}.

Earlier \cite{anisotropy} we concluded that the saturation of the
 magnetoresistance (MR) for Si-MOS samples in parallel field $B_{\rm sat}$
is more related to the
$g^*$-factor of {\em localized} electrons  than that of the conducting ones.
We think the same is true for the magnetoresistance at $B=B_c$.
To illustrate  this conclusion,
we compare in Table II  the $g^*m^*$ values  (i) directly measured
for mobile electrons from Shubnikov-de Haas effect \cite{gm},
(ii) the data which  follow from the slope, $(dn_c/dB_c)$, of the curves in Figs.~2 and 5,
and (iii)  the ones which follow from the slope $(dn/dB_{\rm sat})$ of the
density dependence of the saturation field \cite{anisotropy}.

At $r_s=5.3$, the $g^*m^*$ values determined from
Shubnikov-de Haas oscillations and from MR in parallel field  are comparable.
However, at $r_s=7.6$ they differ
by a factor of two:
$g^*m^*$  determined from the SdH effect
increases with $r_s$ as expected
for conducting electrons \cite{gm}; in contrast,  $(g^*m^*)_{\rm B_c}$ and
$(g^*m^*)_{\rm B_{\rm sat}}$ are close to
the value $2m_b$ ($m_b=0.19$ is the bare band mass)
anticipated for the localized electrons.

\begin{center}
\begin{minipage}{3.2in}
\begin{table}
\caption{Comparison of the $(g^*m^*)$ and $g^*$
measured from Shubnikov-de Haas effect (labeled `SdH')
\protect\cite{gm}
 with values estimated from the MR in parallel field.
Label `$B_c$' denotes the data derived from  $dB_c/dn_c$,
and label `$B_{\rm sat}$' denotes the data from $dB_{\rm sat}/dn$
\protect\cite{anisotropy}.}
\begin{tabular}{|c|c|c|c|}
$r_s$ & $\frac{1}{2}(g^*m^*)_{\rm SdH}$ & $\frac{1}{2}(g^*m^*)_{\rm B_c}$
& $\frac{1}{2}(g^*m^*)_{B_{\rm sat}}$\\
\hline
5.3 & 0.44  & 0.48 & 0.35 \\
7.6 & 0.665 & 0.27& 0.34
\end{tabular}
\label{gm}
\end{table}
\end{minipage}
\end{center}

We close this section with a note that the attempts of describing
$B_{\rm sat}$ and $B_c$
in terms of  the  diverging  $g^*$-factor \cite{shashkin_0007402,vitkalov_divergence},
face with  problems of explaining (a) why  the `divergence' takes
place at progressively (and essentially) lower $r_s$ values (i.e. higher $n_c$ values)
as disorder increases and (b) why the diverging $g^*$ values are different from
those measured in Shubnikov-de Haas effect.
The comparison
presented  in this section illustrates a rather complex behaviour
of the magnetoresistance in the parallel field; definitely, a  thorough
theoretical consideration is needed in order to use this effect for
extracting the electron spin properties.

\subsection{Comparing the effects of $B_{\parallel}$ and temperature
on conduction}
\begin{figure}
\centerline{
\psfig{figure=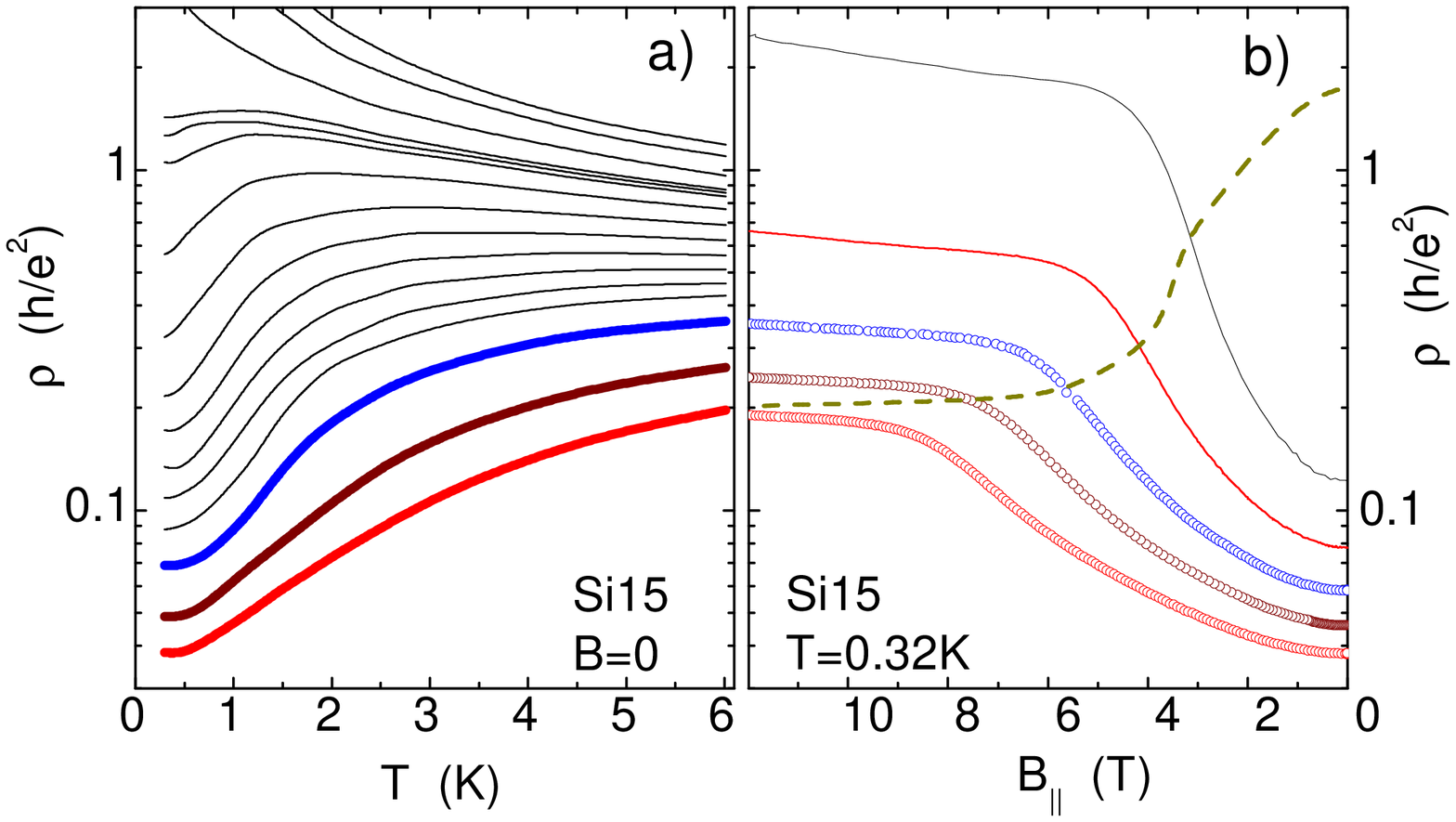,width=230pt,height=170pt}
}
\begin{minipage}{3.2in}
\vspace{0.1in}
\caption{
Resistivity vs temperature (left panel) and parallel field (right panel)
for different densities. Dashed line indicates the boundary for
activated/nonactivated regime. On the left panel the densities are,
from bottom to top:   1.84,  1.62,  1.40, 1.29, 1.23, 1.18,
1.12, 1.07, 1.01, 0.96, 0.91,
0.90, 0.89, 0.85, 0.79, $0.76\times 10^{11}$cm$^{-2}$.}
\label{R(T)_R(B)}
\end{minipage}
\end{figure}

We would like to point to the interesting
similarity between the action of the parallel magnetic field and
the temperature on the resistivity. Figure~\ref{R(T)_R(B)} demonstrates that
for sufficiently high carrier density, $n \gg n_c$, the
application of high magnetic field increases resistivity by about
5-6 times and  restores nearly the same zero-field `Drude'
resistivity value as the temperature does: i.e. $\rho(B > B_{\rm
sat}, T \ll T_F) = \rho(B=0, T \gtrsim T_F )$. The similarity
between the effect of the field and temperature holds only while
conduction remains `non-activated' (cf. Figs.~5).

The dashed line
in Fig.~\ref{R(T)_R(B)}\,b, depicts the boundary (same as in Fig.~4) between
the temperature-activated and `non-activated' transport regimes. When the carrier
density becomes lower than $n_c(B_{\rm max})$ but is still bigger
than $n_c(B=0)$, the parallel field
drives resistivity through this boundary at some field $B_c(n)$.
As a result, the conduction mechanism in magnetic field changes
and the similarity between Figs.~\ref{R(T)_R(B)}\,a and \ref{R(T)_R(B)}\,b breaks.
For this reason, the uppermost curves in Fig.~\ref{R(T)_R(B)}\,b
show much bigger increase in the resistance with field than with
temperature \cite{note}. It is noteworthy, that
 at low density, the magnetoresistance saturation takes place
in the activated regime, at $B_{\rm sat} > B_c$, whereas at higher density the saturation
 occurs in the `non-activated' metallic regime (see Fig.~\ref{B_c&B_sat}.

For the same reason, a very low magnetic field may be sufficient
to drive the conduction from `non-activated' to the `activated' regime
if the density $n$ is chosen slightly less than $n_c$. In other
words, this may be observed when  $\rho(B=0)$ is chosen very close
to but less than $\rho_c(B=0)$  (see  Figs.~5\,a and 8\,b).
In light of this note the  destruction of the metallic
conduction in `arbitrary small field' \cite{simonian97}
and the $(B_{\parallel}/T)$ scaling of the resistivity at high
$\rho$ reported in Ref. \cite{simonian97}
obtain a transparent interpretation \cite{note}.

\subsection{Magnetic field driven disorder}
\vspace{-0.1in}

It was verified experimentally \cite{vitkalov_0101196,anisotropy}
that in Si-MOS structures the parallel field couples to the electron
spins only. Our results, therefore relate the action of the spin
polarization and the increase in disorder. This may occur for
example (i) via weakening of the screening in magnetic field
\cite{dolgo}, (ii) due to inter spin-subband scattering
\cite{murzin_98,pudalov97a,vitkalov_0008456,tutuc}, or
(iii) via increase in the Coulomb scattering caused by
`undressing' (i.e. depopulation and charging) of filled
interface traps in any model of the interface
traps \cite{klap_das,am,kozub,temporal}.

As we demonstrate further,
with the only assumption on the {\em disordering role of the parallel magnetic
field}, we obtain a
qualitative description  of all presented data.
We use this assumption as a phenomenological model;
although we are unable to distinguish which one  (or a combination)
of the microscopic mechanisms is
responsible for the parallel field magnetoresistance, we
evaluate various options in comparison with the data.

The similarity between the $\rho(T)$- and
$\rho(B_{\parallel})$-traces in Fig.~\ref{R(T)_R(B)} seems easy to explain with
the  mechanism of screening which is temperature-
\cite{dsh,stern,gold} and magnetic field-dependent  \cite{dolgo}.
However, comparing this mechanism with  the data we find a number
of principle disagreements:
\begin{itemize}
\item
a) The magnitude of the
magnetoresistance (factor of 5 - 5.5  in Fig.~7   in the
`metallic' range) is larger than the screening mechanism can
provide \cite{dolgo}.
\item
b) At sufficiently high parallel field, $B
> B_{\rm sat}$, magnetoresistance almost saturates; this field was
shown to be very close to the complete spin polarization of 2D
carriers \cite{vitkalov_0101196}. For the screening mechanism, even though
the spin system is polarized, the temperature dependence of resistivity
should be as strong as in zero field.
The dashed line in Fig.~1\,b corresponds to the density $n(B_{\rm sat})$
at which the magnetoresistance for this sample saturates. Obviously,
$\rho(T)$ depends on temperature much weaker than that in Fig.~1\,a for $B=0$.
\item
c) The Hall resistance data \cite{vitkalov_0008456} also does not
support the screening mechanism.
\end{itemize}

Last but not least:  in Ref.~\cite{pud_Hall} a
deviation of the weak field Hall resistance from its classical
value, was observed close to the critical density, particularly,
on the `metallic' side. This means that for $n$ close to $n_c$
there arises an excess or deficit of delocalized carriers; this
observation obviously supports the latter scenario (iii). For this
mechanism to work, the filled traps (or localized carriers, which
is the same for us) should lift in energy with parallel field.
This may take place for example, if the relevant traps form the
upper Hubbard band \cite{kozub}, or if the traps are easy or
spontaneously spin-polarized. We considered the latter
possibility in Ref.~\cite{anisotropy} in order to explain {\em
the disorder dependence of the field for saturation of the
magnetoresistance}. We note that the spin polarization of
localized states is favored by the broken inversion symmetry of
the interface.

Interface defect charges originating from the lack of stoichiometry
are intrinsic to Si/SiO$_2$ system; their
typical density is 10$^{12}$cm$^{-2}$ for a
thermally grown dioxide \cite{hori}.
As parallel field increases,
the  band (or upper band) of localized carriers should lift in energy
and gets `undressed' after passing
through the Fermi energy. This will cause `turning-on'
the charged scatterers  and corresponding
increase in  the scattering rate of the mobile carriers.

The carriers released from the traps will join the conduction
band and may be detected via a deviation in the Hall voltage
as $n$ approaches $n_c(B)$; the latter may be driven either by decrease in $n$ or
by increase in $B_{\parallel}$.
The trapped charge apriory may be of arbitrary sign
which depends on the interface chemistry and
growing processing.
Consequently, the
deviation in the Hall voltage may be of arbitrary sign
as well. Whereas the contributions into the Hall voltage
of the released electrons and holes may compensate each other,
it is not the case for the spatially separated charged
scatterers. The number of `turned-on' scatterers
 $N_i$ is therefore expected to be much bigger than the
deviation in the Hall resistance.
In Ref.~\cite{pud_Hall}, we observed    $dV_H/V_H$
of the order of  $(1-10)$\% in different samples
for the carrier density $n \sim
10^{11}$\,cm$^{-2}$; from this figure a lower estimate for
the  number of `turned-on' scatterers,
$N_i \gg (10^9 - 10^{10})$\,cm$^{-2}$, follows.

It is noteworthy, the inter(spin)-subband scattering
\cite{murzin_98,pudalov97a,vitkalov_0008456,tutuc} is not an
alternative option but may be supplementary to the mechanism
of the magnetic field driven disorder and charged traps. However,
recent experiments \cite{gm,valley} have shown that
the scattering times in different spin and valley-subbands in (100)-Si are close to each other
even though the degree of polarization is $\sim 30$\%. This sets substantial constraints on
those two-band models which imply the intra-subband scattering times to be different.

For completeness, we mention that
the parallel field magnetoresistance in Si-MOS samples
was measured  earlier by Bishop, Dynes and Tsui  \cite{bishop}
and by Burdis and Dean \cite{burdis}
(though as a weaker effect on more disordered samples)
and associated with electron-electron interaction and
 Zeeman splitting \cite{lee&rama}.
In this interpretation, however,  the values of the interaction constant
$F$ were found to be unomalously large and to
{\em decrease as temperature decreases} \cite{burdis} which is not consistent
with theoretical expectations.
We obtained qualitatively similar results on high mobility samples
\cite{aps99,elsewhere}.
In other words, $\rho(B_{\parallel},T)$ in the `metallic'
range {\em does not scale} as a function of $(B/T)$
which is expected for the Zeeman term
$\delta\sigma = - 0.084 (F/2) (g^*\mu_B B/k_BT)^2$
\cite{lee&rama}. It seems therefore unlikely the strong  magnetoresistance
in Si-MOS samples to be  caused by electron-electron interaction.

We limited our consideration by strong changes
in resistivity, $\delta\rho(B)/\rho\sim 1$.
Parallel magnetic field
causes an additional small effect on the quantum corrections to the
conductivity, by decreasing the phase breaking length
\cite{elsewhere}; the latter though does not contradict our concept
of the magnetic field driven disorder.

\vspace{-0.1in}
\section{Implementation of the model for analysis of other data}
In order to verify the above model we compared it
with all available data for Si-MOS sample and found rather good
agreement. For the sake of shortness, we present here the
comparison only with those data for which it was explicitly stated
in Ref. \cite{kravRMP} that ``the enormous response observed at
low temperatures is a consequence of effects other than parallel
magnetic field-induced changes in carrier density or disorder
strength''.

\subsection{Example 1.}
\begin{figure}
\centerline{
\psfig{figure=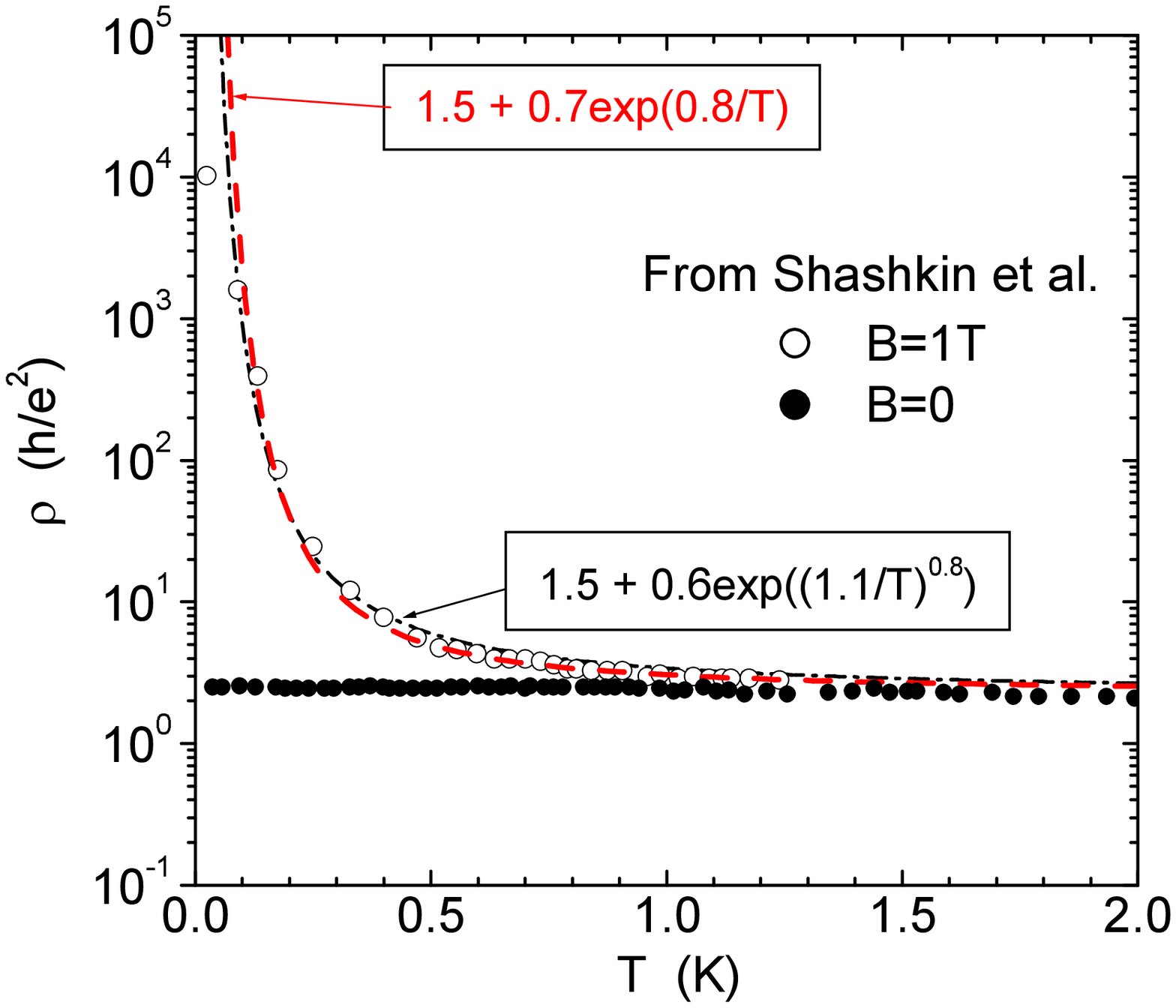,width=240pt}
}
\begin{minipage}{3.2in}
\vspace{0.1in} \caption{ a) Resistivity of Si-MOS sample vs temperature at
zero field and at $B_{\parallel}=1$\,T. Density
$n=0.75\times 10^{11}$cm$^{-2}$. Reproduced from Fig.~9
of Ref. \protect\cite{kravRMP}. Dashed and dash-dotted curves are fitting dependences.}
\label{kravRMP}
\end{minipage}
\end{figure}

\vspace{-0.1in}
We begin with results obtained by Shashkin et al. and reproduce them
from Fig.~9 of Ref.~\cite{kravRMP}.
These data shown in Fig.~\ref{kravRMP}
 ``dramatically demonstrate an extreme sensitivity of the
resistance to parallel field'' at $T=30$\,mK. The localized
behaviour which appears to be absent for the curve at $B=0$ is
restored in a magnetic field'' (cited from Ref. \cite{kravRMP}).
Based on this
contrast it was  stated in Ref.~\cite{kravRMP}
that the effect of the parallel field is
something different from the increase in disorder.

We plot in Fig.~\ref{kravRMP} the model curves using a conventional
temperature activated dependence
\begin{equation}
\rho(T)=\rho_0 + \rho_1\exp(\Delta/T).
\label{activated}
\end{equation}
The effect of the magnetic field in our model is presented in
the magnetic field dependence of the activation energy
$\Delta(B_{\parallel})\equiv
\Delta(n-n_c(B_{\parallel}))$.
We fitted the curves to the  data using three adjustable
parameters $\rho_0$, $\rho_1$ and $\Delta$; their numerical values
are indicated on the figure. Whereas there is a certain correlation between
 $\rho_0$ and $\rho_1$, we focus on the
activation energy $\Delta$ which is well defined and
almost independent of other parameters.

\vspace{0.2in}
\vspace{-0.1in}

The carrier density $n=0.75\times
10^{11}$\,cm$^{-2}$  is set equal to the critical
density,  $n=n_c(B=0)$. By definition, the almost horizontal curve for
$B=0$ corresponds to  $\Delta=0$. Fitting  the upper
curve we obtained $\Delta_{fit} = 0.8$\,K which is reasonably
consistent with a value calculated on the basis of our results
from Fig.~5: $\Delta_{calc}=
(d\Delta/dn)[n-n_c(B=1T)] = 0.6$\,K.
For this estimate we used the coefficient
$d\Delta/dn = (4-5)$\,[K $10^{-11}$ cm$^2$] measured
earlier \cite{Delta} for variety of high mobility samples.
The lowest temperature point deviates substantially from the fitting curve;
we presume this is due to either electron overheating by $\sim 0.07$\,K \cite{akk}
or by a crossover to another $T-$dependence. We also didn't care about a minor
discrepancy for
$T>1.5$\,K; this  may be eliminated by a simple modification of the model as
demonstrated further.

Anyhow, the increase in the measured resistivity by four orders of magnitude
(which  authors of Ref. ~\cite{kravRMP} considered as an extraordinary effect)
is at least 10 times smaller than the increase demonstrated  by the conventional
exponentional dependence Eq.~(1) (dashed curve).
In general, the model curve fits the data rather well and some further improvement
may be obtained by varying the critical index; an example is illustrated in Fig.~8
by the dash-dotted curve and the lower formula.
The success of this fitting, as well as the consistency
between the calculated and fitted  values for $\Delta$ confirm that
the action of the magnetic field is simply described
by the increase in the critical density value $n_c(B)$ related to the  magnetic field
induced disorder.

\subsection{Example 2}
\begin{figure}
\centerline{
\psfig{figure=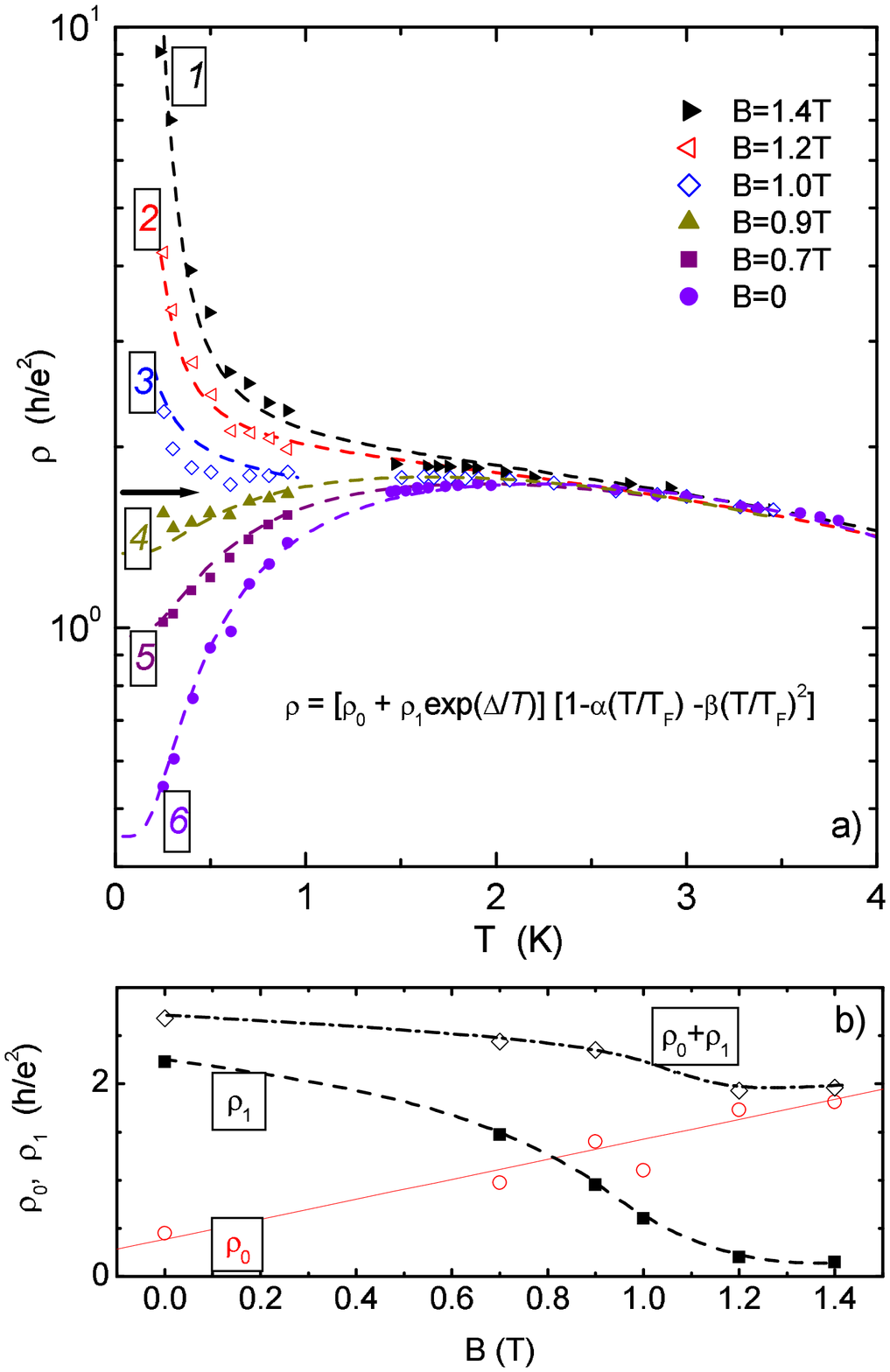,width=240pt}
}
\begin{minipage}{3.2in}
\vspace{0.1in} \caption{ a) Resistivity vs temperature for
different fixed parallel fields (shown at the top). Dashed lines
show fitting curves in the insulating range (curves {\it 1, 2, 3}) and
metallic range (curves {\it 4, 5, 6}).
Horizontal arrow marks an
estimated location of the `critical' trace.
b) Two fitting parameters vs parallel field.
Experimental data are reproduced from Ref.~\protect\cite{simonian97}}
\label{simonian97}
\end{minipage}
\end{figure}

\vspace{-0.1in}
We now turn to the data by Simonian et al. \cite{simonian97}
and reproduce them in  Fig.~\ref{simonian97}\,a;
it shows the temperature dependence of
$\rho$ for Si-MOS sample
 measured in fixed various in-plane fields between 0 and 1.4\,T.
 The data were discussed in Ref. \cite{kravRMP} as a demonstration
 of an abrupt  onset of the metallic behaviour and abrupt
 development of the magnetoresistance.

We note first that the overall $\rho(T)$ behavior for this sample is
different from that shown in Fig.~\ref{kravRMP}: here is no room for a
temperature independent curve
in the range  2\,K$> T> 0.2$\,K and the `separatrix'
$\rho(T, n=n_c)$ is tilted   \cite{akk}.
Correspondingly, the resistivity curves for this sample plotted vs
density, for various temperatures (in the range (0.2-2)\,K) {\em would not
cross each other at a single density} \cite{note_crossing}.

The carrier density $n=0.883\times 10^{11}$\,cm$^{-2}$ in Fig.~\ref{kravRMP}\,a
was set close to $n_c$ so that the effect of the field is large. Initially,
at $B_{\parallel}=0$, the resistivity demonstrates a `metallic'
temperature dependence typical for high mobility Si-MOS samples
(cf. Fig.~1\,a), but as field increases, it transforms to a typical
`insulating' one. We
estimate the transition between `metallic' and `insulating'
behaviour to occur in the region  marked with a horizontal
arrow, $n_c(B=0.93\,T)=n$.
From this number and using the empirical dependence
$n_c(B)$ from Fig.~5 we estimated
the critical density at zero field, $n_c(B=0)=0.76 \times
10^{11}$\,cm$^{-2}$;
this agrees within  5\% with the number (0.802) given in Ref. \cite{simonian97}.

We fitted all curves first with the exponential
dependence
$\rho= \rho_0 + \rho_1\exp(\frac{\Delta}{T})$, same as Eq.~(\ref{activated}).
For the `metallic' region (curves {\it 1, 2, 3}), this
empirical dependence with $\Delta <0$ is
known to describe $\rho(T)$ for different samples
\cite{pudalov97a,akk,app}.
 For the `insulating' region (curves {\it 4, 5, 6}),
Eq.~(\ref{activated}) with $\Delta>0$ has a meaning of the
temperature-activated dependence
(though the same might be done with a variable-range
hopping exponent).
The three adjustable parameters, $\Delta $, $\rho_0$ and $\rho_1$,
where fitted for each curve;
their values are given in
Table ~2. We found the fit is rather good for low temperatures,
$T<2$\,K.

\vspace{-0.1in}

Eq.~(1) can not obviously describe the high temperature
region $T>2$\,K, therefore we
also applied a modified expression
which takes into account the non-exponential `tilt' of the
experimental data:
\begin{equation}
\rho= \left(\rho_0 + \rho_1\exp(\frac{\Delta}{T})\right)
\left(1-\alpha\frac{T}{T_F} -\beta\left(\frac{T}{T_F}\right)^2\right).
\label{tilted}
\end{equation}
The two additional parameters $\alpha = 0.194$ and $\beta= 0.46-0.66$ are
uncorrelated with the other ones, and are directly obtained from the
tilt at high temperatures.

Figure~\ref{simonian97} shows that Eq.~(\ref{tilted})  fits the data
rather well over the whole range of $T$
whereas for low temperatures, $T<2$\,K, both fits,
with Eqs.~(\ref{activated}) and (\ref{tilted}),
are almost indistinguishable. The resulting
parameters for both fits
are given in Tables~2, and 3, correspondingly.
Comparing the fitting curves with data,
we conclude that the ``abrupt changes'' discussed in Ref. \cite{kravRMP}
are nothing more than just a conventional exponential dependence.

The field dependence of the two principle parameters,
$\rho_0$ and $\rho_1$ is shown in Fig.~\ref{simonian97}\,b:
$\rho_0$ increases  with field
demonstrating the disordering role of the field
(though we don't think that the almost linear increase
is a universal feature for different samples).
The strength of the `resistance drop' in the `metallic' range,
$(\rho_0+\rho_1)/\rho_0$, decays with field, also similar to
the decay induced by disorder
\cite{noscaling,mauterndorf}. The critical resistivity, $\rho_c$,
in our model is given by the sum $\rho_0 +\rho_1$;
we find a qualitative similarity
between the two independently calculated curves,
$\rho_c(B)$ in  Fig.~5 and $\rho_0 +\rho_1$ vs $B$ in Fig.~\ref{simonian97}\,b.

\begin{center}
\begin{minipage}{3.2in}
\begin{table}
\caption{Parameters  of the six curves
for the fit with  Eq.~(1).
$n_c$ is in $10^{11}$\,cm$^{-2}$,
$\rho_0, \rho_1$ are in $h/e^2$, and $\Delta$ is in K.}
\begin{tabular}{|c|c|c|c|c|c|c|}
curve & B (T)& $n_c(B)$ &  $\rho_0$ & $\rho_1 $ & $\Delta_{\rm fit}$ & $\Delta_{\rm calc}$ \\
\hline
1 & 1.4 & 0.945 & 1.45   & 0.24 & 1     & 0.31   \\
2 & 1.2 & 0.919 & 1.29   & 0.35 &  0.6  & 0.18  \\
3 & 1.0 & 0.893 &  1.09  & 0.6  & 0.2   & 0.05   \\
4 & 0.9 & 0.881 &  1.0   & 0.8  & -0.15 & -0.01 \\
5 & 0.7 & 0.855 &  0.9   & 1.1  & -0.55 & -0.14\\
6 & 0   & 0.765 &  0.45  & 2    & -0.71 & -0.59
\end{tabular}
\end{table}
\end{minipage}
\end{center}

\begin{center}
\begin{minipage}{3.2in}
\begin{table}
\caption{Parameters of the six  curves shown in Fig.~\ref{simonian97}
for the fit with  Eq.~(2).}
\begin{tabular}{|c|c|c|c|c|}
curve & B (T)&  $\Delta_{\rm fit}$ & $\rho_0$ & $\rho_1 $ \\
\hline
1 & 1.4 &  1.0   & 1.82 & 0.15  \\
2 & 1.2 &  0.6   & 1.76 & 0.2  \\
3 & 1.0 &  0.2   & 1.1  & 0.6 \\
4 & 0.9 &  -0.75 & 1.4  & 0.95 \\
5 & 0.7 &  -0.76 & 0.97 & 1.47 \\
6 & 0   &  -0.75 & 0.45 & 2.23
\end{tabular}
\end{table}
\end{minipage}
\end{center}

For the curve {\it 3}, the fit is not perfect
because it oscillates
vs temperature, which is beyond the frameworks of the
model.
The only steep feature may be found in the field dependence of
$\rho_1$ at the transition from insulating to metallic behavior;
this however may be caused by a minor  mismatching of the oversimplified
model curves in the metallic and insulating ranges.

The `activation energy' $\Delta$ depends on the choice of the
model and on average is 2-3 times larger than the values estimated
from Fig.~5 (cf. the right columns in Table 2).
This discrepancy is presumably caused by the
admixture to the data of a
non-exponential temperature dependence which is essentially strong for the
considered sample  and which can not be separated
at low temperatures from the exponential one. For example, we were
able to achieve a reasonable fit of the data with Eq.~(2),
using the {\em calculated}
in Table~2 values $\Delta_{\rm calc} = (d\Delta/dB)\times(n_c-n)$ when
we modeled the `tilted separatrix' in  Eq.~(\ref{tilted})
by a  power low factor, $T^{-p}$,
rather than by the polynomial factor.

In Ref.~\cite{kravRMP} it was stated that ``the effect of magnetic field cannot
be ascribed solely to a field-induced change in the critical electron
density''.
We verified this possibility and
arrive at the opposite conclusion.
The success of our fitting confirms  that the action of the
parallel field, at least,  to the first approximation consists in progressive
increase of disorder. This is described by
increasing the $T$-independent scattering rate
$\rho_0(B_{\parallel})$, decreasing  the magnitude of the
resistivity drop, $(\rho_1 + \rho_0)/\rho_0$,
and increasing the critical carrier density
$n_c(B_{\parallel})$; the three parameters are not independent and within
the same material system
may be reduced only to the single parameter, e.g. $n_c(B_{\parallel})$.

\section{Summary}
\vspace{-0.1in}
To summarize, the results reported in this paper demonstrate that
the effect of the parallel magnetic field on
resistivity in high mobility Si-MOS samples (though via  Zeeman
coupling) is similar to that of disorder and, to some extent,
to that of temperature. In other words, {\em parallel
magnetic field increases disorder}. The temperature dependence
of resistivity  for
various disorder and  magnetic fields may be reduced to
the dependences of the `critical' density $n_c$ on magnetic field and
on disorder; the  changes in the resistivity drop with temperature,
$(\rho_0+\rho_1)/\rho_0$, caused by  disorder and  magnetic field may
 be also  mapped onto each other.
 We find a similarity between the action of  the
parallel field and temperature for the region of carrier density
and field where conduction remains `non-activated'. These findings set
constraints on the choice of the developed microscopic models.

The analogy between the parallel field and
disorder points to the existence of a
sub-band of localized
carriers.
Such band of localized carriers should lift in energy as a function of the
parallel field, gets `undressed' after passing through the Fermi
energy and cause the increase in the scattering rate.
The strong evidence for this mechanism is the variation in the
Hall voltage observed in the vicinity of the metal-insulator
transition. We can not exclude also the contribution of the
inter spin-subband scattering, as a complementary
feature to the above
mechanism. However, our Shubnikov-de Haas data \cite{gm,valley} indicates that the
mobility in different spin and valley sub-bands of conducting electrons
are almost equal and the intersubband  scattering does not play a major
role in the themperature- and parallel field- dependence of the resistance.
Applying the  model of a field dependent disorder,
we find a qualitative explanation of
the whole set of presented results.

\section{Acknowledgements}
\vspace{-0.1in}
One of the authors (V.P.) acknowledges
hospitality of the Lorenz Centre for theoretical physics at
Leiden,  where part of this  work was done and reported at the
workshop in June, 2000. The work was supported by NSF
DMR-0077825, Austrian Science Fund (FWF P13439), INTAS (99-1070),
RFBR (00-02-17706), the Programs \lq Physics of solid state
nanostructures', \lq Statistical physics', \lq Integration' and
`The State support of the leading scientific schools'.

\end{multicols}


\begin{references}
\bibitem{note}The magnetic field driven transition from non-activated to  activated
conduction regime causes a strong exponential
magnetoresistance $\delta \rho(B)  \propto
\exp(-\Delta(B_{\parallel})/T)$,
where $\Delta(B) \propto (B-B_c)$; therefore, $\rho(B_{\parallel})$ scales as $f((B-B_c)/T)$.
A theoretical consideration see in Ref.~\protect\cite{kozub}.
\bibitem{note_crossing}For this reason, the smeared
curves crossing in Figs.~ 2 and 4 can not be considered
as a proof for the absence of MIT in parallel field.
\end{references}
\end{document}